\documentclass[conference]{IEEEtran}
\IEEEoverridecommandlockouts
\usepackage{fnpct}
\usepackage{cite}
\usepackage{amsmath,amssymb,amsfonts}
\usepackage{algorithmic}
\usepackage{graphicx}
\usepackage{textcomp}
\usepackage[dvipsnames]{xcolor}
\usepackage{enumitem}
\usepackage[disable]{todonotes}
\usepackage{tabularx} 

\usepackage[final]{changes}
\definechangesauthor[name=TW, color=blue]{TW}
\definechangesauthor[name=PS, color=orange]{PS}
\definechangesauthor[name=MB, color=red]{MB}
\definechangesauthor[name=AR, color=ForestGreen]{AR}

\usepackage{url}
\presetkeys%
    {todonotes}%
    {inline,backgroundcolor=yellow}{}

\usepackage[latin1]{inputenc}
\usepackage{tikz}
\usetikzlibrary{shapes,arrows}

\usepackage{array}
\newcommand{\PreserveBackslash}[1]{\let\temp=\\#1\let\\=\temp}
\newcolumntype{C}[1]{>{\PreserveBackslash\centering}p{#1}}
\newcolumntype{R}[1]{>{\PreserveBackslash\raggedleft}p{#1}}
\newcolumntype{L}[1]{>{\PreserveBackslash\raggedright}p{#1}}

\def\BibTeX{{\rm B\kern-.05em{\sc i\kern-.025em b}\kern-.08em
    T\kern-.1667em\lower.7ex\hbox{E}\kern-.125emX}}

\DeclareMathOperator{\strand}{\land\land}

\begin{document}

\newcommand{\task}{\tau} 
\newcommand{\dataout}{y}
\newcommand{\datain}{\theta}
\newcommand{\timestep}{t} 
\newcommand{\node}{\xi}

\newcommand{\TheTitle}{Doubt and Redundancy Kill Soft Errors---Towards
Detection and Correction of Silent Data Corruption in Task-based
Numerical Software}

\title{
  \TheTitle
  \thanks{
    This work has received funding from the European Union's Horizon 2020
    research and innovation programme under the grant agreement No 823844 (project ChEESE).
    Tobias acknowledges the support from the
    ExCALIBUR Phase I grant ExaClaw (EP/V00154X/1) and ExCALIBUR's
    cross-cutting project EX20-9 Exposing Parallelism: Task Parallelism (grant
ESA 10 CDEL).
The Exascale Computing ALgorithms \& Infrastructures Benefiting UK Research
(ExCALIBUR) programme is supported by the UKRI Strategic Priorities Fund. 
The programme is co-delivered by the Met Office on behalf of PSREs and EPSRC on behalf of UKRI partners, NERC, MRC and STFC.  }
}


\author{
 \IEEEauthorblockN{Philipp Samfass}
 \IEEEauthorblockA{
 \textit{Department of Informatics} \\
 \textit{Technical University of Munich}\\
 Garching, Germany \\
 samfass@in.tum.de}
 \and
 \IEEEauthorblockN{Tobias Weinzierl}
 \IEEEauthorblockA{
  \textit{Computer Science} \\
  \textit{Durham University}\\
  Durham, United Kingdom \\
  tobias.weinzierl@durham.ac.uk
 }
 \and
 \IEEEauthorblockN{Anne Reinarz}
 \IEEEauthorblockA{
  \textit{Computer Science} \\
  \textit{Durham University}\\
  Durham, United Kingdom \\
  anne.k.reinarz@durham.ac.uk
 }
 \and
 \IEEEauthorblockN{Michael Bader}
 \IEEEauthorblockA{\textit{Department of Informatics} \\
 \textit{Technical University of Munich}\\
 Garching, Germany \\
 bader@in.tum.de}
}

\maketitle

\begin{abstract}
 Resilient algorithms in high-performance computing are subject to rigorous
non-functional constraints.
Resiliency must not increase the runtime, memory footprint or I/O demands too
significantly.
We propose a task-based soft error detection scheme that
relies on error criteria per task outcome.
\replaced[id=PS]{They}{It} formalise how \added[id=MB]{``dubious'' an outcome is, i.e.\ how likely it contains an error.}
Our whole simulation is replicated once, forming two teams of MPI ranks
that share their task results.
Thus, ideally each team handles only around half of the workload.
If a task yields large error criteria values, i.e.~is dubious, we compute the
task redundantly and compare the outcomes. 
Whenever they disagree, the task result with a lower error likeliness is
accepted.
We obtain a self-healing, resilient algorithm which can compensate silent
floating-point errors without a significant performance, I/O or memory footprint
penalty. 
Case studies however suggest that a careful, domain-specific tailoring of the
error criteria remains essential.

\end{abstract}


\section{Introduction}

%
%
Without a revolutionary hardware re-design, a massive further reduction of clock frequency, 
or an increasing power budget accommodating hardware failure correction, 
we have to assume that exascale machines will fail frequently compared to today's machines \cite{Dongarra:14:ApplMathExascaleComputing,Argonne:2014}.
Empirical data \replaced[id=AR]{leads us to}{make us} expect a linear correlation between the system
size and the failure rate \cite{Schroeder:2010:LargeScale}.
The mean time between failures (MTBF) will shrink.
Simulation codes thus have to improve their resiliency. 
In particular, they have to become able to identify machine errors and to handle them.

%
%
Resilient codes traditionally run through two phases:
First, they spot the errors which can either materialise in machine part failures (hard errors) or wrong results (soft errors).
\replaced[id=AR]{Soft errors are easy to spot if they materialise in exceptional values ($\pm\infty$, NaN, e.g.). In numerical simulations it is tricky to identify them otherwise, as we typically work with approximations.  We have to define how far off from a result is considered to be a soft error, while we might struggle to determine this difference given that we typically do not know the exact solution. }{Soft errors are easy to spot if they materialise in exceptional values ($\pm\infty$, NaN, e.g.), but it is tricky to identify them otherwise, as we typically work with approximations in numerical simulations: we have to define how far off from a result is considered to be a softerror, while we might struggle to determine this difference given that we typically do not know the exact solution. }
The ``simplest'' approach \deleted[id=PS]{to soft error detection}is to not detect \replaced[id=PS]{soft errors}{them} at all, 
but to rely on an iterative algorithm, and iterate 
as long as soft errors continue to pollute the outcome \cite{Bronevetsky:2008, Bronevetsky:2018, Austin:2015,
Shantharam:2012,sdc_multigrid}.
To actually detect errors, codes can rely \deleted[id=PS]{on external triggers to flag errors} 
on algorithmic a-posteriori checks \deleted[id=PS]{(checksum approach)}\cite{Reinarz:2018}, 
or they can run multiple redundant computations and compare their outcomes \cite{Herault:2015,Varela:2010,redMPI}.
The last approach requires codes to run the same calculation on different machines or machine parts: 
at least twice to detect errors or even three times to label the ``correct'' solution via a majority vote.
In a second phase, resilient codes have to fix the wrong data.
Here, three strategies are on the table:
Codes can rely on algorithmic fixes---checksums for example allow them to
reconstruct results, while the aforementioned iterative algorithms have the
correction built in.
They can rely on a rollback to a previous snapshot which has been declared valid,
or they can swap in a redundant data set from a valid, redundant calculation.
All strategies assume that soft errors arise sporadically and the machine remains, in principle, intact.

%
%
In a supercomputing context, these resiliency strategies need to satisfy important non-functional requirements.
A resiliency strategy \textbf{should not introduce} significant 
\begin{enumerate}[label=(\roman*)]
 \item additional synchronisation between otherwise independent calculations:
    synchronisation hinders scalability;
\item additional communication bandwidth or latency:
   network bandwidth and responsiveness are precious resources that quickly develop into a bottleneck if stressed too much;
\item I/O needs:
I/O operations are traditionally by magnitudes slower than compute and communication tasks and thus slow down the calculation;
\item additional memory footprint:
supercomputing codes typically try to increase the memory usage per node already to avoid strong scaling stagnation effects. 
\end{enumerate}

%
%
\noindent
Our work starts from the observation that many simulations decompose
their work into work items (tasks over mesh cells, e.g.) \replaced[id=AR]{which}{and} allow \added[id=AR]{us} to define strong or
weak confidence metrics on the outcome of these items:
Negative mass density or NaNs unambiguously flag wrong data. 
Sudden changes in eigenvalues feeding into an admissible time step size or sudden
oscillations make outcomes dubious---they might be actually correct but the changes might also
stem from a soft error\deleted[id=MB]{---and so forth}.
In the absence of an algorithmic postprocessing step which cures and eliminates
errors (cmp.~limiter-based techniques \cite{Reinarz:2018} in our case), we
propose to run our task-based simulation twice in parallel.
As long as tasks yield outcomes where we are confident that they are reasonable,
we make the two replicas share their outcomes.
Effectively, each replica only computes half of the work and relies on its
counterpart to compute the remainder \cite{samfassISC}.
As soon as a task outcome is dubious, our code waits for the counterpart
simulation to compute the outcome redundantly, and we then compare the results.

%

%
%
Soft errors, i.e.~silent data corruption, are notoriously difficult to spot if
there is no strong and immediate validation, such as a hash value that reports data corruption reliably and
immediately after the error has arisen.
In this case, only comparisons to redundant results with majority vote help to
identify and fix them.
Such an approach is infeasible in HPC, as it synchronises,
triples the compute workload, and requires resources to host checkpoints.
Due to task-based error \replaced[id=PS]{criteria}{indicators}, we can  
offer soft error detection without any immediate synchronisation\replaced[id=AR]{. We}{, since we} work on a per-task level, asynchronously, and run the bit-wise
redundancy checks only upon demand.
The error \replaced[id=PS]{criteria}{indicators} also allow us to replace a majority vote with a
confidence vote, while result sharing among trusted outcomes reduces the cost
of the redundant computations.
In many cases, silent data corruption can immediately be corrected, while the
algorithm also provides evidence when rollback-and-recompute becomes mandatory.

\added[id=PS]{Our strategy towards saving the cost of replication by sharing 
outcomes is different from other work which runs replicas at a reduced execution rate for power savings~\cite{lazy_shadowing}.
}
\added[id=PS]{In contrast to task replay techniques in asynchronous many-task runtime systems~\cite{amt_resilience}, our approach has reduced algorithmic latency---the replicated computation runs
in parallel---yet has a higher memory footprint to store results temporarily
until they are approved.
It is also able to recover from hard errors (not shown), as we work
with real rank redundancy.}

%
%
Our concepts are illustrated within the wave equation solver ExaHyPE
\cite{cpc_exahype} which uses an explicit time stepping scheme to tackle the
underlying hyperbolic equations
\added[id=PS]{and which relies on an independent library called teaMPI \cite{samfassISC} for transparent replication of ranks and for task outcome sharing between those ranks.}
Explicit time stepping schemes are notoriously difficult to equip with
resiliency, since they typically operate with large data sets and are computationally demanding, while
the absence of an iterative solver step or diffusion implies that errors spread
out, propagate and pollute the outcome if they are not immediately corrected.
Since we neither need a tailored error correction scheme
\cite{Reinarz:2018}, nor checkpointing, nor multiple redundant computations
\cite{redMPI}, our ideas improve significantly upon the state-of-the-art
how to deliver resilient simulation codes, plus they are of broad
applicability and relevance:
As long as a solver's algorithmics can be
broken down into small tasks and credibility metrics can be defined, our
algorithmic ideas are applicable.

%
%
The paper is organised as follows:
We present our key ideas and terminology in Section \ref{section:algorithm} as an abstract
algorithmic framework.
In Section~\ref{section:ExaHyPE}, we introduce our wave equation solver and
highlight how the framework's concept of error \replaced[id=PS]{criteria}{indicators} is realised for our
application.
This potential impact on a wide range of applications is supported by numerical
results (Section~\ref{sec:evaluation}) which we present after Section~\ref{section:implementation}'s
discussion of implementation details.
With a realisation sketch \added[id=PS]{and experimental results} at hand, we can classify and contextualise our
approach and thus highlight its broader impact (Section~\ref{section:evaluation-and-classification}).
A brief summary and outlook in Section~\ref{section:conclusion-and-outlook} close the discussion.

\section{Algorithmic framework}
\label{section:algorithm}

Let an algorithm consist of tasks $\task_i$ that take
some input $\datain _i$ and yield output $\dataout_i=\task_i(\datain_i)$.
Some tasks have temporal dependencies, i.e.
$\task_i \sqsubset \task_j$:
$\task _i$ feeds into $\task _j$ and thus has to 
terminate before $\task _j$ starts.
A task scheduler exploits the freedom for any task pair $\task_i, \task_j: \task_i
\not \sqsubset \task_j \wedge \task_j \not \sqsubset \task_i$ to deploy such tasks $\task_i$ and
$\task_j$ concurrently to the available compute cores as the tasks are
\emph{independent}.

A \emph{hard error} within a task makes our system crash or prevents the task
from terminating.
A timeout can detect the latter situation so hard errors can be reliably detected.
We focus on \emph{silent errors} which make a task yield $\tilde{\dataout_i}$ instead of the correct $\dataout_i^{\text{(correct)}} =
\task_i(\datain_i)$.
As our numerical simulations work with floating point approximations where the
outcome of a task is not absolutely deterministic due to register data transfers
or operation reordering, we characterise a silent error through $|\tilde
\dataout_i-\dataout_i^{\text{(correct)}}|>tol _\dataout$  ($|.|$ being a suitable norm).
\replaced[id=PS]{For simulations,}{It is in the nature of a simulation that} we do not know $\dataout_i^{\text{(correct)}} $ for a task
prior to its execution and thus struggle to characterise a soft error unless an
error yields NaNs or unreasonable data.
The latter term remains to be defined.

Let a \emph{team} be the set of MPI ranks that are used for a program run.
If we run a simulation completely redundantly, we have two teams $A$ and $B$.
Both issue the same tasks $\task_i^A$ and $\task_i^B$.
The two team schedulers however might deliver different task execution sequences
for tasks without a total order.
As long as both teams yield correct data, i.e.~do not suffer from soft errors, 
$|\dataout_i^A - \dataout_i^B| \leq tol_\dataout$. 
$tol_\dataout$ depends on the machine precision and error propagation during 
the task, so usually encodes a relative error in IEEE floating point
precision with a fixed number of significant bits.

An \emph{error criterion} is a hash function $f \colon \dataout_i \mapsto
[0,\infty]$, where
$f(\dataout_i) \leq tol_{f}$ indicates that there is no reason to assume that a soft
error has occurred during the calculation.
$f(\dataout_i)=\infty $ signals that something went wrong.
Any value in-between highlights that a silent error might have crept into the
result.
$f$ thus quantifies to which degree the task outcome is dubious.

Due to the numeric nature, there is however no guarantee that
$f(y_i)=0$ implies that no error has occurred and there is no direct correlation
between $tol_\dataout$ and $tol_f$.
\added[id=PS]{$f(\dataout_i^A) > f(\dataout_i^B)$ does not necessarily mean that $\dataout_i^A$ is wrong. 
However, $f(\dataout_i^A) \gg f(\dataout_i^B)$ suggests that $|\dataout_i^A-\dataout_i^{\text{(correct)}}|>|\dataout_i^B-\dataout_i^{\text{(correct)}}|$. 
As silent errors are very unlikely to affect both outcomes, the one with the lower $f$-value is likely correct.}
Depending on the context, $f$ evaluates an absolute value of $\dataout_i$ or takes
historical data such as previous values of a numerical solution into account
and calibrates $\dataout_i$ accordingly.

\added{We combine multiple error criteria $f_k$ to an \emph{error indicator} $\phi(y_i)$,
which can yield boolean or numerical values.
When using booleans, we define $\phi(y_i)$ via  a logical predicate, such as $\exists f_k : f_k(\dataout_i)>tol_{f_k}$.
When using numerical values, we may again use a tolerance $\phi(y_i) > tol_{\phi} $ to indicate dubious results. 
}
\replaced[id=PS]{Different error criteria might exist for different tasks.}{Different error indicator functions might exist for different tasks and there might
be multiple error indicator functions per task outcome.}
To streamline our notation, we focus on one task type $\task $ only.
Let $f_1, f_2, \ldots$ then denote different criteria for
task type $\tau$, with tolerances $tol _1, tol _2, \ldots$.


Our algorithmic framework relies on few key ingredients:
\begin{itemize}
  \item There are two teams $A$ and $B$. Their scheduler is initialised with
  a bias seed: If two tasks $\task_i$ and $\task_j$ 
  are independent and if $A$'s scheduler gives $\task_i$ a high priority and $\task_j$ a low priority, then $B$'s
  scheduler gives $\task_j$ a high priority and $\task_i$ a low one. This bias manifests
  non-deterministically in different task completion sequences.
  \item Whenever a task completes, we can compute its error \replaced[id=PS]{criteria}{indicators} $f_k$.
  \item Each team has a \emph{local cache} of task outcomes; and hence of their
  error \replaced[id=PS]{criterion}{indicator} values.
  We use this cache to inform the other team whenever we have completed a task
  earlier than the team counterpart, and we use the cache also to temporarily
  store local task results if we are not sure whether they have been corrupted.
\end{itemize}

The algorithm then reads as follows (cf. Figure~\ref{fig:control_flow_correction_and_detection}):
\begin{enumerate}
  \item If a task is to be executed and the task outcome is not yet in the
  cache, compute it locally.
  \item If a local task computation yields error \replaced[id=PS]{criterion values}{indicator hashes} which indicate that
  the result is trustworthy, share the task outcome with the counterpart in the
  other team. Continue with the computation.
  \item If a local task computation yields error \replaced[id=PS]{criterion values}{indicator hashes} which indicate that
  the result might be compromised, share this outcome, too, but place the local
  result in the local cache for the time being. Do not accept the task outcome
  yet. Let the computing core progress with another task\added[id=PS]{ and check the outcome later}.
  \item If a task is to be executed, if the task outcome is already in the local
  cache, i.e.~has been sent in by the other team, and if this incoming data has
  error \replaced[id=PS]{criteria}{indicators} which suggest that these data are trustworthy, skip the
  local computation and accept the data sent in as task result.
  \item If a task is to be executed, the task outcome is in the local cache and
  is flagged as dubious, run the calculation locally \added[id=PS]{and share the outcome. Check the local outcome if it is not trustworthy.}
\end{enumerate}

\noindent
The scheme is complemented by some garbage collection that discards task results that
are no longer required.
This happens for tasks that drop in as problematic while the local computation
yielded a result with low error criteria values, or two task outcomes that cross in the
network.

\tikzstyle{decision} = [diamond, draw, fill=blue!20, 
    text width=2.5cm, minimum width=3cm, minimum height=1cm, text centered, node distance=3cm, inner sep=0pt, aspect=2]
\tikzstyle{block} = [rectangle, draw, fill=blue!20, 
    text width=5em, text centered, rounded corners, minimum height=4em]
\tikzstyle{line} = [draw, -latex']
\tikzstyle{cloudgreen} = [draw, ellipse,fill=green!20, node distance=3cm,
    minimum height=2em]
\tikzstyle{cloudred} = [draw, ellipse,fill=red!20, node distance=3cm,
    minimum height=2em]
\tikzstyle{cloudorange} = [draw, ellipse,fill=orange!20, node distance=3cm,
    minimum height=2em]

\begin{figure}
\centering
\begin{tikzpicture}[node distance=3cm, auto, every node/.style={scale=0.5}]

\node [cloudgreen] (init) {start};

\node [decision, below of=init] (outcomeavailable) {task outcome $y_i'$ available};
\path [line] (init) -- (outcomeavailable);

\node [decision, below right of=outcomeavailable] (hasComputed2) {\hspace{2em}hasComputed($y_i$)\hspace{2em}};
\path [line] (outcomeavailable) -| node[anchor=south]{no} (hasComputed2);
\path [line] (hasComputed2) -| node[yshift=3.5ex]{yes} (outcomeavailable);
\node [block, below right of=hasComputed2] (compute2) {compute $y_i$};
\path [line] (hasComputed2) -| node[anchor=south]{no} (compute2);
\node [block, below of=compute2] (share) {share $y_i$ with other team};
\path [line] (compute2) -- (share);
\node [decision, below of=share] (local_dubious) {isDubious($y_i$)};
\path [line] (share) -- (local_dubious);
\node [cloudgreen, below of=local_dubious,yshift=0.88cm] (completed3) {$y_i$ valid};
\path [line] (local_dubious)-- node[anchor=west]{no} (completed3);
\node[right of=local_dubious](anchor){};  
\path [line] (local_dubious) -- (3,-6.64) node[yshift=2ex, anchor=west]{yes}   |-(init);

\node [decision, below left of=outcomeavailable,xshift=-2cm] (hasComputed) {\hspace{2em}hasComputed($y_i$)\hspace{2em}};
\path [line] (outcomeavailable) -|  node[anchor=south]{yes} (hasComputed);

\node [decision, below left of=hasComputed,xshift=-1cm] (checkRemote) {$|y_i-y_i'|\leq tol_y$};
\path [line] (hasComputed) -| node[anchor=south]{yes} (checkRemote);
\node [cloudgreen, below left of=checkRemote] (completed1) {$y_i$ valid};
\path [line] (checkRemote) -| node[anchor=south]{yes} (completed1);
\node [decision, below right of=checkRemote,xshift=-0.5cm] (checkErrInds) {$y_i'$ more likely};
\path [line] (checkRemote) -| node[anchor=south]{no} (checkErrInds);
\node [block, below left of=checkErrInds,yshift=0.5cm] (copy2) {copy $y_i'$ into $y_i$: $y_i=y_i'$};
\path [line] (checkErrInds) -| node[anchor=south]{yes} (copy2);
\node [cloudgreen, below of=copy2] (completed6) {$y_i$ valid};
\path [line] (copy2) --  (completed6);

\node [decision, below of=checkErrInds,xshift=0.1cm] (equalErrInds) {$\forall k : f_k(y_i)==f_k(y_i')$};
\path [line] (checkErrInds) -- node{no} (equalErrInds);
\node [cloudorange, below of=equalErrInds] (moderate) {keep $y_i$};
\path [line] (equalErrInds) -- node{yes} (moderate);
\node [cloudgreen, below right of=equalErrInds] (completed5) {$y_i$ valid};
\path [line] (equalErrInds) -| node[yshift=-5ex]{no} (completed5);

\node [decision, below right of=hasComputed,xshift=1.5cm] (remote_dubious) {isDubious($y_i'$)};
\path [line] (hasComputed) -| node[anchor=south]{no}  (remote_dubious);
\path [line] (remote_dubious) --  node{yes} (compute2);

\node [block, below left of=remote_dubious,xshift=0.3cm] (copy)  (copy) {copy $y_i'$ into $y_i$: $y_i=y_i'$};
\path [line] (remote_dubious) -| node[anchor=south]{no} (copy);
\node [cloudgreen, below of=copy] (completed4) {$y_i$ valid};
\path [line] (copy) -- (completed4);

\end{tikzpicture}
\caption{Control flow to compute, share, check and correct the outcomes $y_i$ of a task execution. $y_i'$ denotes a task outcome from another team.}
\label{fig:control_flow_correction_and_detection}
\end{figure}
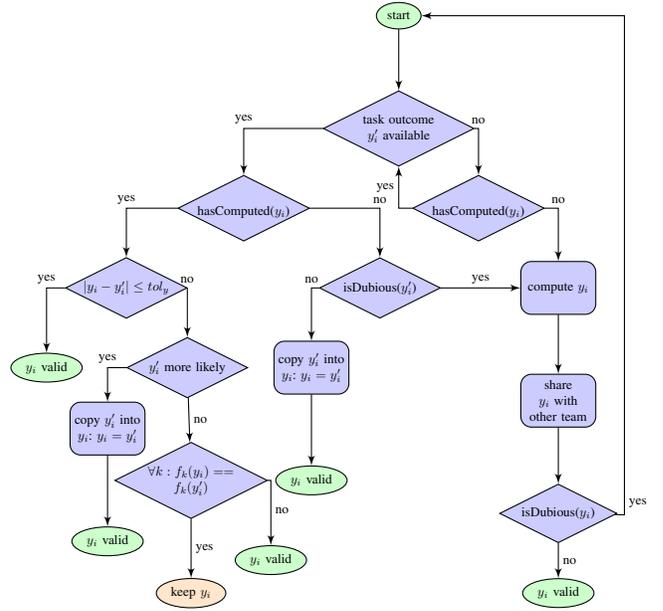


As long as all tasks yield trustworthy results, the above scheme with the biased
scheduler shares the actual compute workload equally among the two teams.
As soon as a dubious task outcome \deleted[id=PS]{ with a high error indicator value}arises on either team, the
task result is backed up in the local cache.
We then have to match two outcomes for the same task within the cache\added[id=PS]{ based upon the error criteria}:

\begin{itemize}
  \item If one task outcome has high error criteria values and the other task
  outcome has low error criteria values, we assume that the latter is the valid
  result and that the former includes some silent error. We accept and continue
  with the valid data.
  \item If both task outcomes have \replaced[id=PS]{the same error criteria values and $|y_i^A-y_i^B|\leq tol_y$, our algorithm had become over-dubious: It turned out that the result is ``kind of surprising''
  but valid.}{a high error indicator value, our algorithm had
  become over-dubious: It turned out that the result is ``kind of surprising''
  but likely valid.}
  \item \added[id=PS]{If both task outcomes have the same error criteria values but disagree (i.e., $|y_i^A-y_i^B|> tol_y$), we have spotted a silent error and need to moderate (e.g., keep the local $y_i$).}
  \item If both task outcomes have error criteria values of $\infty$, we have spotted a
  hard or intrinsic algorithmic error. The code has to terminate immediately,
  and it is up to the user to decide whether to alter the setup or to restart
  from the latest checkpoint.
\end{itemize}

\todo{Is this ok here? Or do we need to write about the combination of multiple values?}

\noindent
It is straightforward to supplement the algorithmic framework with some
timeouts:  
If the redundant task's outcome does not arrive within a certain timeframe
for dubious local results, or if one team sends out task outcomes but does
not receive any outcomes from the other over a longer period, it is reasonable
to assume that the counterpart team suffers from a hard error.
In this case, the healthy team can still checkpoint. 
Checkpoint-restart thus integrates seamlessly into our algorithmic framework.
As hard errors are out-of-scope in the present work, we neglect further timeout
discussions from hereon.

\section{Demonstrator and benchmark scenario}
\label{section:ExaHyPE}

\begin{figure}
 \centering
  \includegraphics[width=0.8\linewidth]{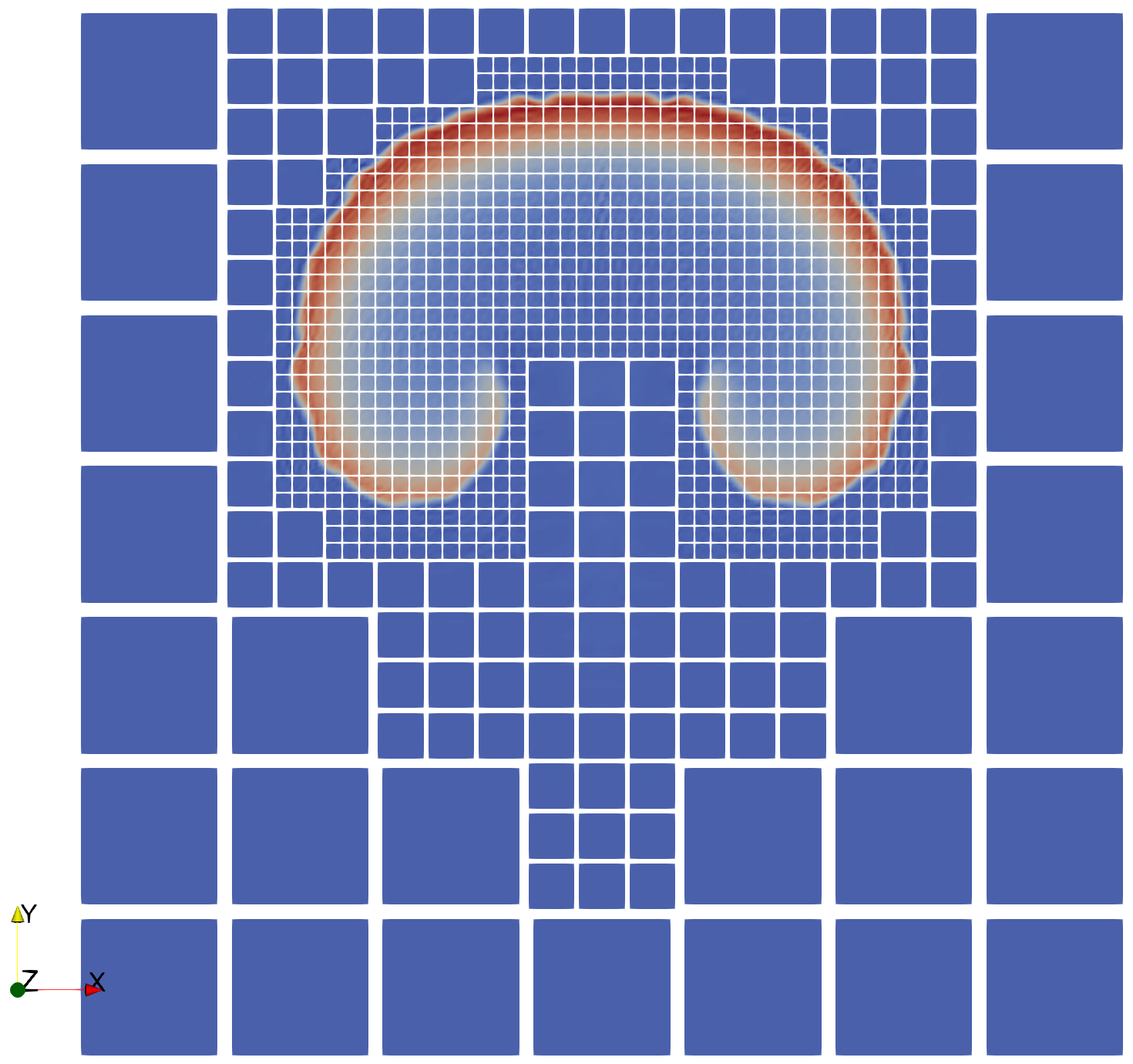}
  \caption{
    Two-dimensional cut through our three-dimensional benchmark setup:
    We solve the compressible Navier Stokes
    equations to simulate a rising cloud on a dynamically adaptive
    Cartesian grid \cite{Weinzierl:19:Peano}.
    \label{fig:cloud}
  }
\end{figure}

We realise our algorithmic ideas within the ExaHyPE engine \cite{cpc_exahype},
relying on previous work to share task outcomes between teams using the teaMPI library \cite{samfassISC}.
ExaHyPE provides a parallel software infrastructure to solve hyperbolic partial differential equations (PDEs) in their first-order form, e.g.:
\begin{equation}
\frac{\partial Q }{\partial \timestep} + \nabla \cdot F(Q, \nabla Q) = S(Q).
\label{eq:generic}
\end{equation}

\noindent
We solve for a vector of unknowns $Q(x,\timestep)$ in time and space. 
The formulation requires a flux tensor $F(Q)$
and a source term $S(Q)$ plus possibly further terms omitted here.
ExaHyPE's numerics are based upon the ADER-DG scheme \cite{Dumbser:2006}, which
discretizes the PDE per element in space and in time with high order polynomials
and develops the solution in time with a predictor-corrector scheme.
The computational domain is discretised by a dynamically adaptive Cartesian
mesh \cite{Weinzierl:19:Peano}.
Each mesh cell (cube) carries a solution polynomial.
Per time-step, we run through three sub-steps:
\begin{enumerate}
 \item A cell-local solve yields the so-called \emph{space-time predictor}
 $\hat Q(x,\timestep)$, a high-order polynomial approximation \replaced[id=PS]{of}{to} the solution
 within the given cell. This solution neglects the influence of neighbour
 cells. As we compute it per mesh cube, the space-time predictors can be
 executed independently over the mesh.
 \item A Riemann solve at each cube face in space and time computes
 the numerical fluxes across the faces which are induced by $\hat Q$, and thus
 captures the influence of adjacent cells upon each other.
 \item A corrector step combines the space-time predictor $\hat Q$ with the
 result from the Riemann solves into a new solution $Q(x,\timestep+\Delta \timestep)$.
 Corrections are again computed per cell and hence can run independently.
\end{enumerate}

\noindent
Our code combines non-overlapping domain decomposition along the
Peano space-filling curve with task-based parallelisation.
The domain decomposition is realised via MPI.
After each space-time prediction step, we have to exchange the domain boundary
data to allow all ranks to compute all relevant Riemann problems.
We have one data exchange per time step which materialises in one MPI message
burst.
Further data exchange such as a global reduction to determine an admissible time
step size is negligible w.r.t.\ bandwidth.

The tasking classifies all cells per rank into cells that are adjacent to mesh
resolution \added[id=PS]{transitions} or to the MPI boundary and cells that can reside within the rank's
domain interior and thus can be handled with lower priority.
This enclave tasking \cite{Charrier2020} implies that MPI data transfer and
space-time predictor computations overlap, and that we obtain good per-node
scaling \cite{Charrier:19:EnergyAndDeepMemory}.
The space-time prediction is the computationally dominant task type in ExaHyPE  
\cite{Charrier:19:EnergyAndDeepMemory}.
We thus focus on this task $\task $ in the context of our resiliency work.

teaMPI~\cite{samfassISC} is a wrapper around MPI which plugs into MPI's tools interface PMPI.
The teaMPI wrapper hides that we operate with two replicated
teams, by using split communicators such that ExaHyPE is unaware of the
redundancy.
\replaced[id=PS]{ExaHyPE leverages teaMPI's interface for task outcome caching, for communication of task outcomes and for querying whether an outcome is available through a replica computation.}{It also owns the task caching and collects global statistics such as 
late sender patterns. 
However, additional teaMPI interfaces allow ExaHyPE's task backend to
query whether a task outcome is already available through a replica computation.}

In our replica/teaMPI mindset, task outcome caches synchronise with their
MPI rank counterpart only, as both teams $A$ and $B$ employ exactly the same
non-overlapping domain decomposition.
If a task $\task ^A_i$ is spawned on rank $r$ within team $A$, the same task
will be spawned by rank $r$ in team $B$ eventually.
No data exchange between different rank numbers within the two teams is
necessary.
%

\subsection{Test setup}

Our test benchmark solves
the three-dimensional compressible Navier Stokes equations 
\begin{equation}\label{eq:navier-stokes}
 \frac{\partial}{\partial \timestep} 
 \underbrace{
 \begin{pmatrix}
   \rho\\
   \rho v \\
   \rho E
 \end{pmatrix}}_{=Q}
 + \nabla \cdot F(Q, \nabla Q)
=
  \underbrace{
 \begin{pmatrix}
 0\\
 -g k \rho\\
 0
 \end{pmatrix}}_{=S(Q)},
\end{equation}
where
\[
F(Q, \nabla Q) = \begin{pmatrix}
 \rho v\\
 v \otimes \rho v + p I + \sigma\\
 v \cdot \left(  (\rho E + p)I + \sigma \right)
  - \kappa \nabla T
 \end{pmatrix}.
\]

\noindent
Here, $\rho$ denotes the density, $\rho v$ the momentum and $\rho E$ the energy density.
The pressure is given by $p$ and the temperature by $T$.
The constant $\kappa$ is used to model diffusion of the temperature in the term $\kappa \nabla T$.
A stress tensor $\sigma = \sigma(Q, \nabla Q)$ accounts for viscosity. In the source term, $k$ denotes the unit vector in vertical direction and $g$ the gravity of Earth.

\deleted[id=MB]{%
This system employs a flux $F(Q,\nabla Q)$ which introduces diffusive terms and
thus is not covered by
(\ref{eq:generic}).
The resulting system describing the compressible Navier-Stokes equations is not
first-order anymore or hyperbolic, yet can be solved directly with ADER-DG
nevertheless, as long as tiny time step sizes are used and the friction terms
remain very small.
}


To solve the PDE system \eqref{eq:navier-stokes}, we use the ExaHyPE-based ADER-DG solver by Krenz et al.~\cite{krenzcloud} to simulate a rising warm air bubble (see Fig.~\ref{fig:cloud}).
This scenario reproduces the setup from \cite{kelly}, where a perturbation in a
potential temperature field propagates over a background state that is in hydrostatic balance.


\subsection{Error criteria}
ExaHyPE realises an explicit time stepping scheme where the space-time predictor
$\dataout_i = \task_i(\datain_i)$ for each cell maps a polynomial onto an extrapolated polynomial.
Any error that is introduced to the solution perturbs the outcome polynomial.
We may assume that localised errors affecting individual sample points within
our Gauss-Legendre ansatz decrease the smoothness of the overall polynomial.
At the same time, (\ref{eq:navier-stokes}) prevents certain solution values such
as negative densities:


\paragraph{Arithmetic corruption}
If $\dataout_i = \task(\datain_i)$ contains NaNs, the data $\dataout_i$ has been compromised.
We can employ a simple checksum/reduction to identify these cases.
Let $f_{\text{NaN}}(\dataout_i) = \infty$ if $\dataout _i$ holds NaNs.
In this case, the task outcome is invalid.
Otherwise, $f_{\text{NaN}}(\dataout_i) = 0$.
We do not employ a tolerance $tol_{\text{NaN}}$ in this case or may work with
any $0 < tol_{\text{NaN}}$.

\paragraph{Physical corruption}
Our application setup solves conservation laws subject to certain plausibility
checks.
We call them physical admissibility (PA) checks.
 The solver tracks the density and potential temperature of the solution as
 primary variables over the mesh.
 Both are subject to the PDE and have to be non-negative.
 The physical admissibility criterion furthermore can derive a pressure from all
 of the primary quantities plus some material parameters.
 This pressure serves as further
 admissibility criterion, as it always has to be positive, too.
Let $f_{\text{PA}}(\dataout_i) = \infty$ if $\dataout_i$ violates the physical
admissibility check.
Otherwise, $f_{\text{PA}}(\dataout_i) = 0$.
Like $f_{\text{NaN}}(\dataout_i) $, $f_{\text{PA}}(\dataout_i) $ is a boolean
label, too.

\paragraph{Dubious time step size changes}
ExaHyPE relies on adaptive time stepping, i.e.~the time step size $\Delta\timestep$ is not
prescribed, but depends on the largest eigenvalue of the flux (in the above example, both the largest eigenvalue of the flux and the viscous flux), the polynomial order and the mesh size.
It is chosen thus to fulfil the CFL condition, i.e.~follows the speed information spreads through the grid.

We store the time step size \replaced[id=MB]{$\Delta\timestep _i$}{$\timestep _i$} per cell and thus can determine
how significant the time step size per cell changes from one time step to the other.
If we employ global adaptive time stepping, the overall time step size results
from a reduction over cell-local time step sizes.
For the resiliency strategy, solely the local data are of interest.

Let $f_{\Delta \timestep}(\dataout_i) = |\Delta
\timestep^{\text{new}}_i-\Delta \timestep^{\text{old}}_i|/\Delta
\timestep^{\text{old}}_i$.
$f_{\Delta \timestep}(\dataout_i)$ accepts that the time step size per cell changes---as waves enter
or leave the cell---but doubts the task result if this change is, relative to
the previous time step, significant.
The criterion indirectly relates the eigenvalues of the PDE over a cell to historic data.

\paragraph{Solution smoothness evolution}
$\frac{\partial^2}{\partial x_d^2}$ denotes the second partial derivative
operator in direction $d$.
As we work with element-wise polynomial solutions in ADER-DG, it is straightforward to
determine the second derivatives over the sample points $\node_n$ per cell.
Let
\[
f_{\text{Der},d}(\dataout_i) = 
\frac{1}{N}
\sum_{ \node_n} 
\frac{
\left|\frac{\partial^2}{\partial x_d^2}
\dataout_i^{\text{new}}(\node_n)-\frac{\partial^2}{\partial x_d^2}
\dataout_i^{\text{old}}(\node_n)\right| 
}{ 
\left| \frac{\partial^2}{\partial x_d^2} \dataout_i^{\text{old}}(\node_n)
\right|
}.
\]

\noindent
This error metric computes to which extent a newly
computed task outcome increases the maximum second derivative compared to its previous value.
It considers all directions separately.
We evaluate the second derivative at each sample point $\node_n$ 
of the polynomial's Lagrangian representation, and
sum up all obtained values per direction to obtain a single
reduced value $f_{\text{Der}}(\dataout_i):=\sum_d f_{\text{Der},d}(\dataout_i)$.
While small changes of the maximum second derivative are natural as waves propagate, 
very large values of $f_{\text{Der}}(\dataout_i)$ are suspicious.
They flag drastic changes of the solution smoothness.
This can happen for non-linear equations due to wave stiffening, yet is rare.

\paragraph{Combining multiple error criteria}
\added[id=TW]{
A free choice of $tol_{\Delta \timestep}$ and
$tol_{\text{Der}}$ allows the user to incorporate domain-specific knowledge (``is the solution
usually smooth or are shocks/steep gradients typical'' for example)
and facilitates a balancing between sensitivity and speed.
}

\deleted[id=TW]{
Our approach needs to decide whether an outcome is dubious based upon all 
error indicators $f_{\text{NaN}}, f_{\text{PA}}, f_{\Delta \timestep}$ and
$f_{\text{Der}}$
(\deleted[id=TW]{cmp.~}Figure~\ref{fig:control_flow_correction_and_detection}).
Arithmetic corruption and physical corruption are binary choices.
If they flag up ($f_{\text{NaN}}=\infty$ or $f_{\text{PA}}=\infty$), the
outcome of a task is corrupted and hence trivially dubious, too.
}

\deleted[id=TW]{
For the other dubiosity checks, dubiosity is a continuous value.
We leave the choices of $tol_{\Delta \timestep}$ and
$tol_{\text{Der}}$ to the user.
This allows the user to incorporate domain-specific knowledge (``is the solution
usually smooth or are shocks/steep gradients typical'' for example)
and facilitates a balancing between sensitivity and speed.
}

\added[id=PS]{We support rigorous or lazy evaluation of the error criteria.
In the rigorous variant, all error criteria are evaluated. 
If any criterion is violated, the outcome is seen as dubious and needs to be checked further.
We obtain $\phi(\dataout_i) = (f_{\text{NaN}}(\dataout_i)>0) \,\lor\, (f_{\text{PA}}(\dataout_i)>0) \,\lor\, (f_{\text{Der}}(\dataout_i)>tol_{\text{Der}})\,\lor\,(f_{\Delta t}(\dataout_i)>tol_{\Delta t})$ as a boolean dubiosity error indicator.
Here, $\lor$ denotes a strict logical OR.
It ensures that all error indicator values are available for comparisons with a matching task outcome from another team for $\phi(\dataout_i)=1$.  
In the lazy variant, we first evaluate the computationally cheap error criteria $f_{\text{NaN}}, f_{\text{PA}}$ and $f_{\Delta \timestep}$.
Only if one of these pre-filtering criteria is violated, the derivative error criterion $f_{\text{Der}}$ is evaluated. 
The task outcome is dubious if and only if it violates one of the pre-filtering criteria \emph{and} the derivatives criterion, i.e.\ $\phi(\dataout_i) = [(f_{\text{NaN}}(\dataout_i)>0) \,\lor\, (f_{\text{PA}}(\dataout_i)>0) \,\lor\,(f_{\Delta t}(\dataout_i)>tol_{\Delta t})] \strand (f_{\text{Der}}(\dataout_i)>tol_{\text{Der}})$. 
Here, $\strand$ denotes a \emph{non-strict} logical AND.
}
\added[id=TW]{If the pre-filtering criteria are non-dubious, we formally assume
 $f_{\text{Der}}=0$.
}

\added[id=TW]{
 Once two task outcomes' criteria differ, we have to decide which outcome is
 erroneous.}
\added[id=PS]{We base this decision on comparisons of the individual error criterion values, i.e., $y_i^A$ is more likely than $y_i^B$ according to $f_k$ if $f_k(y_i^A)<f_k(y_i^B)$. 
We cascade these comparisons according to the order $f_{\text{NaN}},f_{\text{PA}},f_{\text{DER}}$ and $f_{\Delta t}$ non-strictly, i.e., we stop as soon as one $f_k$ has flagged $y_i$ as more likely. 
This results in a prioritized evaluation.
Criteria checked earlier are not allowed to contradict subsequent criteria for marking an outcome $y_i^A$ as more likely, i.e., we demand that $f_k(y_i^A)==f_k(y_i^B)$ for these criteria.
}
\replaced[id=PS]{This can result in situations in which we cannot decide upon the erroneous outcome.
In these cases, we keep the local outcome and emit a warning indicating that a silent error could not be corrected. 
Further error recovery measures could then be activated to avoid error propagation.}{
 This can lead to divergence, where both replica are flagged as dubious and both
 either keep their data or both swap.
 In practice, we have not observed such a phenomenon.}
 {}

\section{Implementation}
\label{section:implementation}

\subsection{Selection of tasks}

The space-time predictor tasks are reasonable to hook in an error indicator, as
they match the non-functional requirements that we identified for the overall
resilient algorithm:

%
%
(i) The cost to compute a task of interest has to be high compared to
the evaluation cost of the \replaced[id=PS]{error criteria}{dubiosity scalars}.
An upper threshold for the $f$-cost is the computational time to compute
the core task $\task $ itself---%
otherwise, the effective cost per task doubles, and even if the two teams  
perfectly share their outcomes, the absolute time-to-solution remains invariant.
In practice, we expect to see at least some runtime savings in return for
investing twice as many resources and the cost of the $f$-evaluations thus has
to be significantly smaller.
For our space-time predictor, we observe that the cost for the dubiosity checks
are significantly smaller than the $\task $ computation, as the space-time
predictor solves a dense non-linear problem.

%
%
(ii) The number of ready tasks that can be shared between teams has to be high. 
Only if the runtime is dominated by phases when a lot of ready tasks
linger in the system, we can shuffle their execution order and hence profit
from task outcome sharing, and exchange task outcomes while other tasks are
still computed.
For ADER-DG, the first phase per time step, which issues embarrassingly parallel
space-time prediction tasks only, is a perfect fit to this requirement.
Previous work of ours has demonstrated that we can shuffle the execution order
of these tasks slightly by assigning task priorities, and obtain
reasonable task sharing ratios as long as the tasks remain uncorrupted and we
can assume that all incoming task outcomes are valid \cite{samfassISC}.

%
%
(iii) Tasks must, on an academic notion of the task concept, be atomic and final:
\replaced[id=MB]{They must not have any}{They must be isolated, i.e., a task may not have any}
immediate side effects, and it must be possible to delay re-using \replaced[id=MB]{outcomes}{its outcome} in
follow-up computations.
Furthermore, tasks are not allowed to interrupt or spawn further tasks.
Each STP task can run independently to other STP tasks as it accesses only element-local data.
An STP's result feeds into Riemann solves at the adjacent cell-faces of an
element for computing the numerical flux from and to its neighbours, but it does
not directly yield further tasks.
Instead, we wait for the corrections to be finished before we issue the next
type of tasks (Riemann solves).

%
%
(iv) The memory footprint of a task's output must be relatively small.
We have to share task outcomes between rank pairs from different teams, and we
have to cache task results locally whenever a task result drops in or our local
computations suggests that some dubiosity and consistency checks become
necessary.
Furthermore, only small footprints ensure that we can transfer output quickly via
MPI and the task result sharing does not introduce interconnect congestion.
Our preliminary work \cite{samfassISC}, which we use as code base here, has
demonstrated that task outcomes can be shared in a timely manner as long as we
take special care regarding the MPI progression.
However, the present approach still runs risk to double the effective memory
footprint per rank.

%
%
(v) The likelihood that silent data corruption affects the task outcome directly
has to be high. Our approach relies on local checks and immediate
correction of corrupted task outcomes.
STP tasks account for most of the consumed CPU time during an ExaHyPE
run \cite{Reinarz:2018,Charrier:19:EnergyAndDeepMemory}, making them very likely
to be affected by silent data corruption.
\todo{I don't see data that shows a 91\% rate but more around 60\%}
\added[id=TW]{
 Previous work using a simplified oscillation analysis (min/max
 condition) in combination with the physical admissibility criterion
 furthermore suggests that these two criteria can identify up to 60\% of the
 significant silent data corruptions within STPs for the Euler and Einstein
 equations \cite{Reinarz:2018}.
}
However, it remains to be validated experimentally to which degree
\replaced[id=TW]{these insights carry over to our lazy dubiosity
checks, apply to our application domain, and how a selection of dubiosity
tolerances affects both the runtime}{data corruptions can be detected by the dubiosity checks introduced so far }\added[id=PS]{ and the error detection rates}.


While the space-time predicator tasks $\task $ are responsible for the bulk
of the compute cost, they are not the only tasks within our system.
Other, cheap tasks feed into the space-time predictor tasks or follow them,
i.e.~pass their result into predictor tasks of the subsequent time step.
If errors affect these cheaper tasks that are not subject to our team checks,
they will pollute follow-up space-time predictors\replaced[id=PS]{ to which our error detection and correction approach is applied again.}{.
That is, we will not be able to correct the corrupted tasks immediately, but the
algorithm will fix all subsequently polluted tasks afterwards and thus recover
the simulation's state.}




\subsection{Asynchronous checking of task outcomes}

The performance of our approach relies on a highly asynchronous implementation
where teams are not running in lock step mode.
They are not tightly synchronised.
Therefore, a team is never allowed to block for receiving a
redundantly computed result.
Our implementation regularly checks for incoming task
outcomes and receives them whenever available.
 Such a mechanism handles ``unexpected'' messages carrying task outcomes from
 tasks that have been executed earlier on the replicated rank than on the local
 rank.
 Polling the MPI subsystem prevents overflow of the MPI buffers in use 
 and that MPI has to switch to a rendezvous protocol.

Once a task has been computed locally on a team $A$ and its outcome is
considered as dubious, there is no guarantee that
team $B$ delivers a
matching outcome in a timely fashion
(right branch of first check in Figure~\ref{fig:control_flow_correction_and_detection}).
Latency or contention delay any message delivery further.
 Busy waiting for the ``control computation'' therefore is not an option. 

\replaced[id=TW]{
 Whenever we wait for a replicated task's outcome, we want to switch to further
 computations while we wait.
 To be independent from modern MPI+X callback mechanisms~\cite{continuations} that support tasks
 with ``interrupts'' or listeners for incoming MPI messages, we introduce an additional 
 task type which checks and corrects%
}{
 We enforce switching to further computations while waiting for a
 replicated task's outcome by introducing an for checking and correcting 
}
a computed dubious task result.
This task is spawned for each dubious STP task that cannot be checked immediately.
It is created with low-priority and re-schedules itself until a redundantly computed result has been inserted into the local cache.
Then, the consistency checks are performed and, if necessary, the
task outcome is corrected.
The tasks that reschedule themselves logically realise a polling mechanism, 
but the actual polling is spread out over further calculations as further tasks
slot in.


\subsection{Error model}
\label{sec:error_model}

%
%
Our experiments rely on a manual error injection to
facilitate controlled studies.
The underlying error model assumes that silent data corruption happens
exclusively within the STP.
It takes the STP's outcome $\hat Q$, and introduces errors by
adding $\delta Q$ to \replaced[id=PS]{this outcome}{the solution}.
We assume that no other errors arise, and that the silent data corruption
exclusively affects the outcome of floating point calculations.
This is reasonable, as data corruptions on integer data typically lead to wrong
memory accesses or wrong execution logic, such that they materialise almost
immediately in a hard error.

%
%
As our $Q$ in (\ref{eq:generic}) is represented by polynomials in a 
Lagrangian formalism over the cells, 
and as our error injection picks a\deleted[id=PS]{an arbitrary} sample point $ \node_n $ and adds
a value, a silent error alters the per-cell representation of a task outcome.
In our task formalism, we obtain \added[id=TW]{altered Lagrangian weights}
$\tilde {y}_i = y_i+e = \task (\theta _i) + e$,
which translates to a flawed space-time prediction $\hat Q(x,\timestep) + \delta
Q(x,\timestep)$.
An injected error \deleted[id=MB]{in one sample point $ \node_n $} \added[id=MB]{thus} alters the \deleted[id=MB]{total}
representation of the (predicted) solution in the \added[id=MB]{entire} cell, 
\replaced[id=MB]{but}{as we work with a higher-order formalism.
As we employ a Discontinuous Galerkin scheme, it} does not immediately propagate
globally\deleted[id=TW]{:
A flawed STP outcome \deleted[id=MB]{however} feeds into the Riemann problems on the cell's face
and thus effectively travels one cell per time step; in line with the CFL
condition.
\added[id=PS]{The flawed STP, however, manifests itself as an absolute error of
magnitude $|\delta Q|$ at one sample point $\xi_n$ in a single cell locally}}.


\replaced[id=TW]{
Less than 20\% of random bitflips
within a given floating point number actually introduce non-negligible errors in
$Q(x,t)$ for our application \cite{Reinarz:2018}.
We therefore refrain from injecting an error into the task calculation or into
$\theta _i$.
Instead, we fix a value $|e|$ and artificially add this value to one sample
point $\node_n $ of one of the STP outcomes in one time step.
This yields a guaranteed permutation $\delta Q$.
We pick the error location
randomly, i.e.~a randomly chosen cell and a randomly chosen coefficient in its
STP is affected.
We also pick the affected time step randomly.
There is only one single ``bitflip'' which manifests in a significant error,
i.e.~falls into the 20\% category, and this ``bitflip'' is non-persistent,
i.e.~occurs only once.
}{
Previous studies have shown that only a small
fraction of random bitflips ($<20\%$) 
introduce non-negligible errors in the task outcome $Q$.
To facilitate controlled studies and to avoid having to generate
a huge amount of samples, we set the error in our runs
to an error value $e$, testing multiple of
these values in a chosen error range.
We, however, choose the error location
randomly, i.e.~a randomly chosen cell and a randomly chosen coefficient in its
STP is affected.
If not stated differently, we only inject a single error per run, as we gather
our statistics for many very short runs (in the order of $2--5s$ of elapsed wall time to completion).
}
 

\section{Experimental results}
\label{sec:evaluation}


We run all our tests on the SuperMUC-NG\footnote{\url{https://doku.lrz.de/display/PUBLIC/SuperMUC-NG}} supercomputer hosted by the Leibniz-Rechenzentrum in Garching. 
Each SuperMUC-NG compute node features two Intel Xeon Platinum 8174 CPUs (``Skylake'' architecture) with 96 GBytes of main memory and $24$ cores per CPU.
Nodes are inter-connected in a fat tree topology with Intel Omnipath.
We compile and run ExaHyPE and teaMPI with the \added[id=TW]{2019
generation of the} Intel compiler, Intel TBB and Intel MPI\deleted[id=TW]{ for
which we use the 2019 generation of these tools and libraries}.

\subsection{Sensitivity analysis}

We first analyse sensitivity and correctnesss of the error \replaced[id=PS]{criteria}{indicators}.
\added[id=MB]{We systematically prescribe different sizes of errors $e$ 
(in contrast to Section~\ref{sec:error_model} discriminating between positive and negative errors), 
and then conduct 100 test runs for each fixed error size, randomly inserting that error once during the run.}
We track whether the injected error was detected and successfully corrected during the run, 
and thus compute the sensitivity rate, i.e.~
the total number of \emph{corrected runs} divided by the total number of runs.
In the case of a successfully corrected error, the solution remains unaffected by any corruption.
Injected but undetected errors or errors where the
algorithm picks the wrong team as valid propagate and
pollute the long-term wave field.
We run all following parameter studies on a single node with two teams and with a single MPI rank per team. 
We report results only for numerical order $7$ for brevity, although we obtained comparable results at other orders, too.

\begin{figure}
 \centering
  \includegraphics[width=\linewidth]{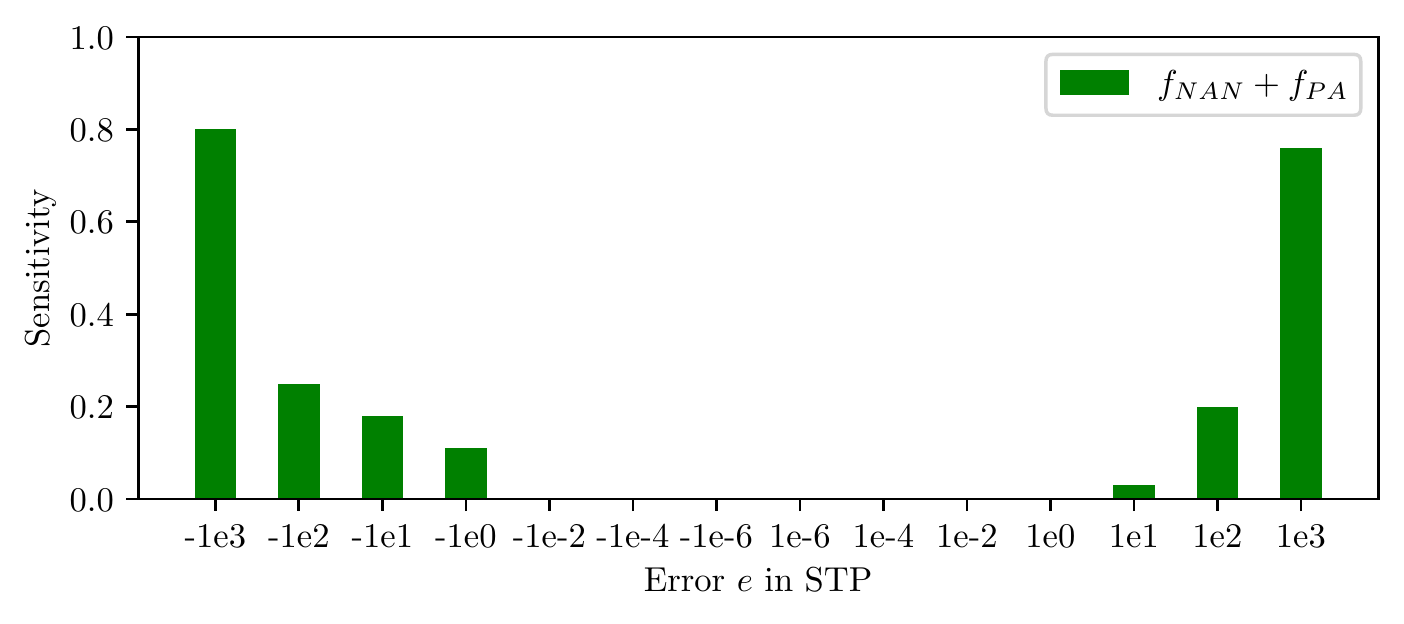}\\
  \includegraphics[width=\linewidth]{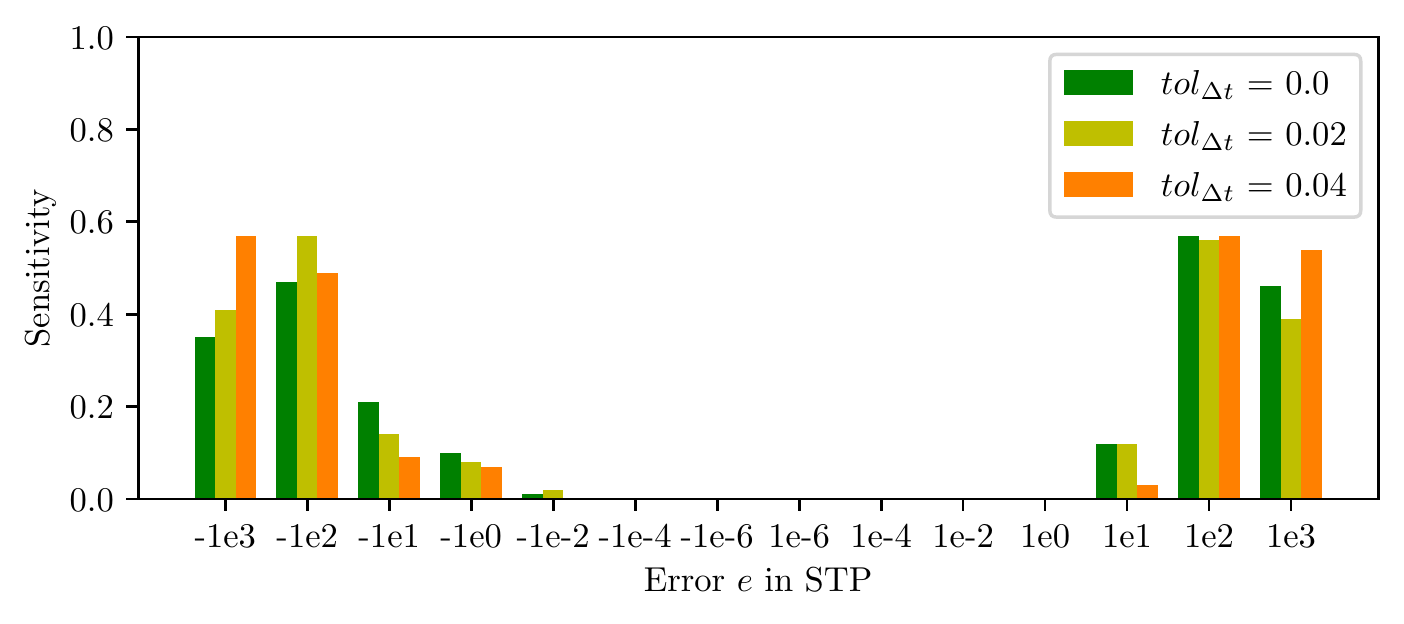}
  \caption{ 
   Top: Error sensitivity for $f^{\text{PA}}+f^{\text{NaN}}$.
Bottom:  Error sensitivity for the time step sizes criterion $f_{\Delta t}$.
\label{fig:sensitivity_nan_pad}
  }
\end{figure}

%
%
\replaced[id=TW]{
For the physical admissibility checks combined with the NaN search,
}{
The results in Figure~\ref{fig:sensitivity_nan_pad} show that
} the error size
determines the sensitivity (Figure~\ref{fig:sensitivity_nan_pad}).
\replaced[id=TW]{The larger an error the more reliably it is}{Typically,
errors of large magnitude can reliably} detected and corrected\deleted[id=TW]{
using these criteria alone}.
\added[id=TW]{
 The NaN criterion is particularly robust, i.e.~finds all NaN in the output
 (not shown), 
 while the combined sensitivity is biased towards negative error contributions.
}

%
%
\replaced[id=TW]{
 As our admissibility criterion searches for negative pressures or }
{As the admissibility checks are also sensitive to} negative density values,
\added[id=TW]{its} sensitivity is higher for negative error
\replaced[id=TW]{contributions}{values} compared to positive values.
\replaced[id=TW]{
 If the random error introduces a negative density in
 one sample point, it is clear that we have an error.
 However, also positive changes of any unknown can violate the admissibility:
 If the density is increased relative to the energy, the pressure
 reconstruction yields a negative pressure.
 The derived quantity harms the physical admissibility.
}{ Not all errors can be corrected: only if a randomly injected error
 affects a density value and renders it negative, the admissibility check can become effective.
}

With the criterion $f_{\Delta t}$, our algorithm similarly reacts mostly to larger error magnitudes with sensitivity values of up to $50\%$ (Figure~\ref{fig:sensitivity_nan_pad}).
The sensitivity rate does not strongly correlate to the tolerance $tol_{\Delta
t}$\added[id=TW]{ and yields solely qualitative metrics, i.e.~dubious
vs.~reasonable}.
\replaced[id=TW]{
 Most smaller solution changes do not alter the time step size even if we 
 compare the time step sizes bit-wisely, i.e.~pick $tol _{\Delta t}=0$.
 This property results from the fact that we use the maximum eigenvalue of the
 result to determine the admissible time step size.
 It is only a change that feeds into the maximum eigenvalue that also triggers
 the error \replaced[id=PS]{criterion}{indicator}.
 While $f^{\text{PAD}}+f^{\text{NaN}}$ seems to yield stronger qualitative
 sensitivity statements for large errors, $f_{\Delta t}$ is more sensitive
 for error in the order of $|e| \approx 10^2$.
}{ 
 Closer inspection of our data reveals that most
 errors do not affect the admissible time step size due to the random choice of error locations which do not necessarily change maximum wave speeds.
If they do, however, the associated error \replaced[id=PS]{criterion}{indicator} values $f_{\Delta t}$ typically exceed even the largest error \replaced[id=PS]{criterion}{indicator} threshold $tol_{\Delta t}=0.04$ and they can reliably be detected.
}

\begin{figure}
 \centering
 \includegraphics[width=\linewidth]{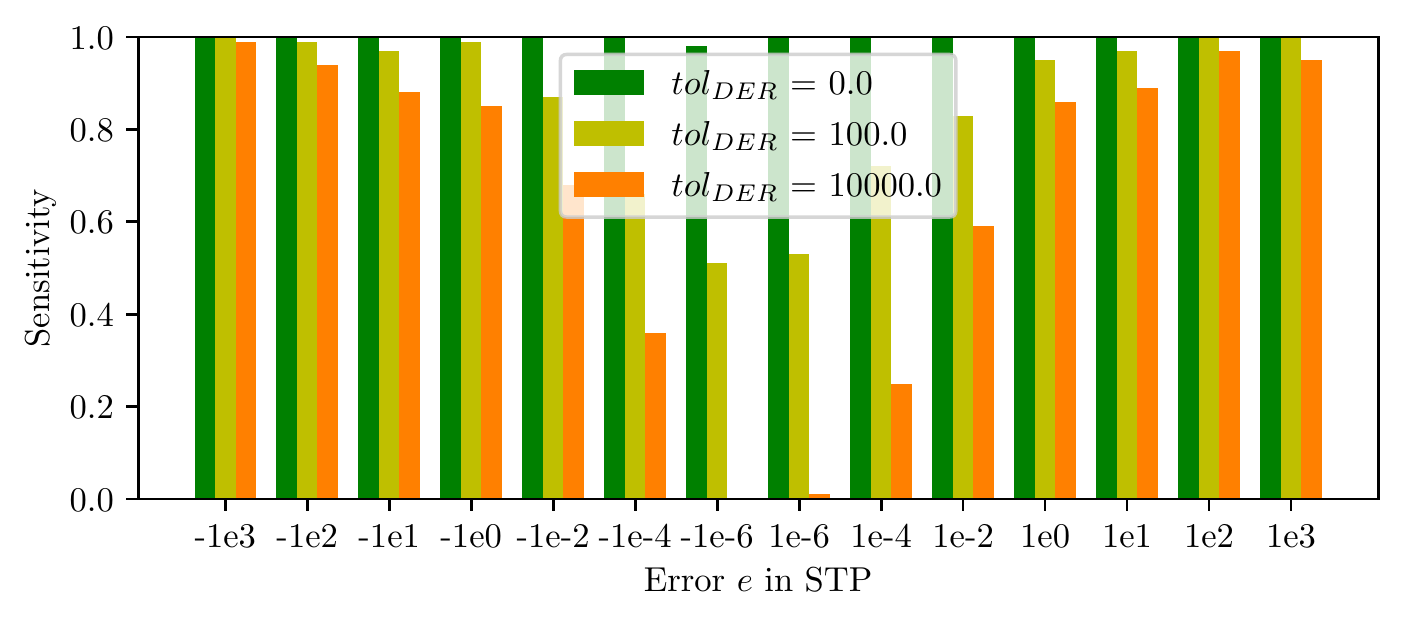}
 \caption{
  Error sensitivity for the derivatives criterion $f_{\text{Der}}$.
  \label{fig:sensitivity_derivatives}
 }
\end{figure}

The derivatives criterion \replaced[id=TW]{is}{turns out to be} the most
sensitive one \added[id=TW]{(Figure~\ref{fig:sensitivity_derivatives})}:
we can \added[id=TW]{spot and} correct errors with high sensitivity of $>0.8$
in most cases.
Sensitivity increases with lower tolerance $tol_\text{Der}$ and---like the other criteria---with larger error magnitudes.

\subsection{Combined dubiosity checks}

\begin{table}
 \caption{
  Average error sensitivities for different configurations of error
  criteria (rounded to two decimals).
  \label{tab:combined_criteria}
 }
\begin{tabular}{ll|rrrp{1.1cm}p{1.1cm}}
$tol_{\Delta t}$ & $tol_{\text{Der}}$ & $f_{\text{PA}}+f_{\text{NaN}}$ & $f_{\Delta t}$ & $f_{\text{Der}}$ & all criteria (rig.) & all criteria (lazy) \\ \hline
0                & 0                  & 0.17                           & 0.16           & 1.00             & 1.00                  & 1.00                       \\
0                & 100                & 0.17                           & 0.16           & 0.86             & 1.00                  & 0.83                       \\
0                & 10000              & 0.17                           & 0.16           & 0.66             & 1.00                  & 0.66                       \\
0.02             & 0                  & 0.17                           & 0.16           & 1.00             & 1.00                  & 0.51                       \\
0.02             & 100                & 0.17                           & 0.16           & 0.86             & 0.87                  & 0.46                       \\
0.02             & 10000              & 0.17                          & 0.16           & 0.66             & 0.77                  & 0.37                      
\end{tabular}
\end{table}

%
%
We next investigate the combination of multiple criteria
\replaced[id=TW]{
 and compare averaged sensitivities
}{
. We average the sensitivity values of the different error values $e \in
\{-1e3,-1e2,...,1e2,1e3\}$ to obtain a single characteristic sensitivity value
per configuration $c$.
In Table~\ref{tab:combined_criteria}, we compare these values 
}
for experiments using only one of the error \replaced[id=PS]{criterion}{indicator} functions $f \in
\{f^{\text{PA}}+f^{\text{NaN}},f_{\Delta t}, f_{\text{der}}\}$ with experiments
using either the rigorous or the lazy combination of \emph{all} presented
error \replaced[id=PS]{criteria}{indicators} \added[id=TW]{(Table~\ref{tab:combined_criteria})}.
In all cases but one\deleted[id=TW]{(first row, where the maximum error
sensitivity of $1$ is already reached with $f_{\text{der}}$)}, the sensitivity
for the rigorous combination of all criteria \replaced[id=TW]{is higher
than the individual ones}{surpasses their individual sensitivities}.
The lazy \replaced[id=TW]{combination yields a reduced
sensitivity}{variants come at reduced
sensitivities, as the highly sensitive error indicator  $f_{\text{Der}}$ is evaluated less aggressively}.

%
%
\replaced[id=TW]{
 The lazy scheme skips some $f_{\text{Der}}$ evaluations which mark a task
 outcome as dubious in the rigorous counterpart.
 It misses out on some dubious results.
 Yet, a combination of different criteria allows an additional calibration of
 the overall code's sensitivity beyond the tuning of tolerances, and thus
 yields more sensitive algorithmics for both the rigorous and lazy
 evaluation.
 While a maximum sensitivity of $1.0$ can be obtained with the tolerances
 $tol_{\Delta t}=0$ and $tol_{\text{Der}}=0$, a rigorous combination of the
 error \replaced[id=PS]{criteria}{indicators} comes at a performance price.
}{
The key takeaway at this point is that error correction can be calibrated to meet the user's needs of confidence for the computed results.
As the error indicators are dependent on the simulated scenario and the properties of the solved PDE, an automated calibration phase is required for determining the sensitivities of different configurations and tolerances.
Short runs can be used for such a calibration. 
While the maximum sensitivity of $1.0$ can be obtained with the tolerances $tol_{\Delta t}=0$ and $tol\_{\text{der}}=0$ and rigorous combination of all error indicators, it also comes at a performance price as discussed followingly.
}

\subsection{Performance}

\begin{figure}[htb]
 \centering
 \includegraphics[width=\linewidth]{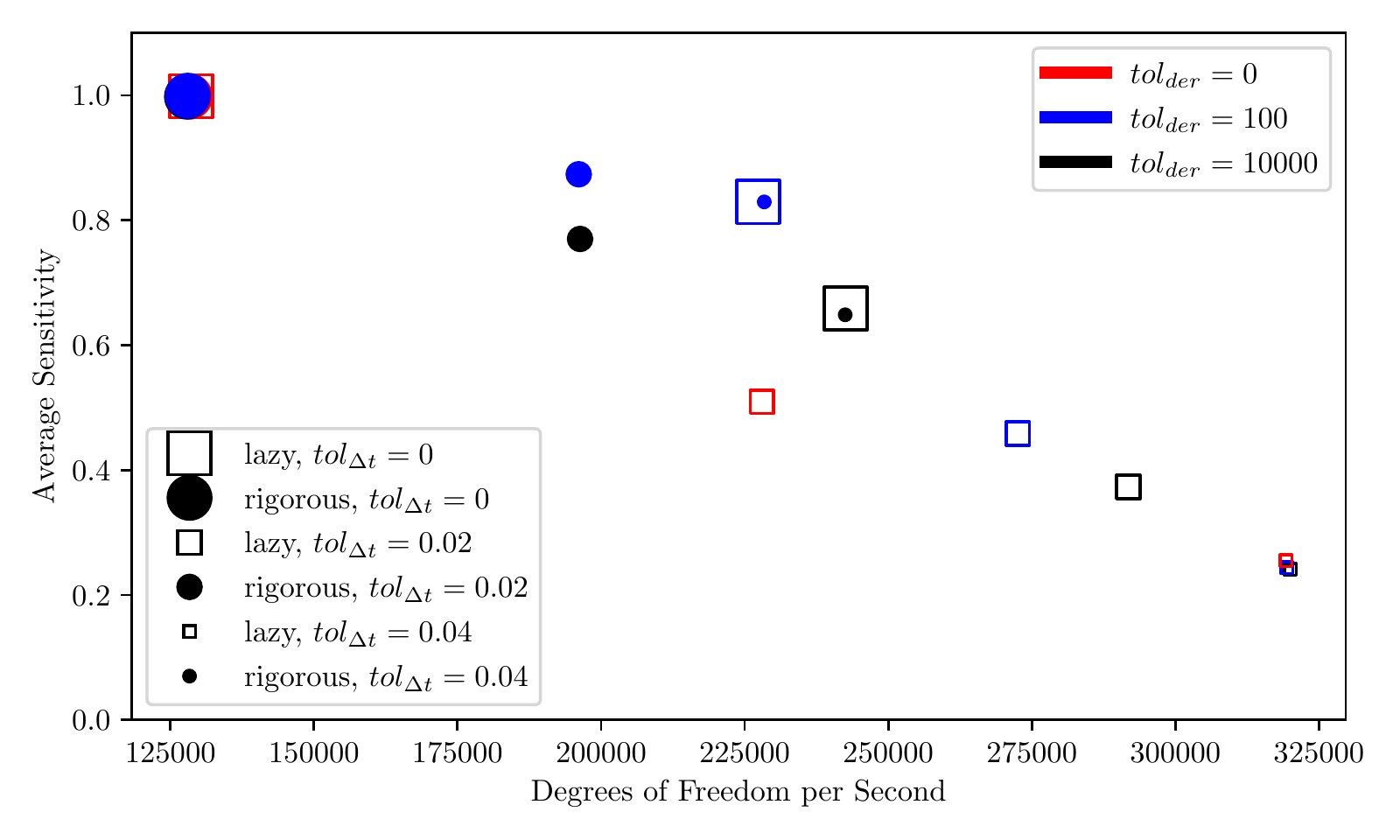}
 \caption{
   Performance/sensitivity tradeoff of different configurations\added[id=TW]{ in a
   two rank setup \added[id=PS]{(single injected error per run)}. \added[id=PS]{The rigorous configurations with $tol_{\text{der}}=0$ (red) are overlapped by others in the upper left corner.}}
   \label{fig:performance_tradeoff}
 }
\end{figure}

\deleted[id=TW]{We next investigate the performance and the overhead of our
approach.}
Both the choice of the error \replaced[id=PS]{criterion}{indicator} functions as well as the \replaced[id=PS]{respective thresholds}{error indicator
thresholds} are influential parameters
as they determine how many task outcomes need to be
validated\replaced[id=TW]{, i.e.~how many (additional) indicator evaluations we
have to run (Figure~\ref{fig:performance_tradeoff}). 
}{
We measured the performance for different configurations for a two-rank setup on a single node, where we ensured that in each run, a single error was injected and corrected.
Figure~\ref{fig:performance_tradeoff} relates the performance for each
configuration to its average sensitivity (cmp. Table~\ref{tab:combined_criteria}).}%
\replaced[id=TW]{%
 An optimal sensitivity of 1.0 (all error are recognised and fixed) can be 
 obtained for many configurations where we evaluate $f_{\text{Der}}$ always.
 However, the throughput is about a factor of three worse than the throughput of
 a code without any sensitivity check.
 This throughput statements refers to a single team run, i.e.~we neglect savings
 due to the sharing of outcomes in a replicated world.
}{
Highest possible sensitivity of about $1.0$ is obtained for many configurations, but it comes at the price of lowest performance, as all computed task outcomes need to be validated.
}
On the other hand, the \added[id=MB]{setup with the highest performance} only exhibits an
average sensitivity of $\approx 0.2$.
We observe a trade-off between performance and sensitivity.
Configurations with lazy evaluation typically come at a penalty on sensitivity compared to their rigorous counterparts, but they achieve higher performance.
Configurations with $tol_{\text{Der}}=100$ (e.g., with lazy evaluation and $tol_{\Delta t}=0$) give the best performance-accuracy trade-off: they achieve a sensitivity of around $0.8$ while coming at a performance that is faster than fully redundant computation.

\subsection{Upscaling}

\begin{figure}
\includegraphics[width=0.98\linewidth]{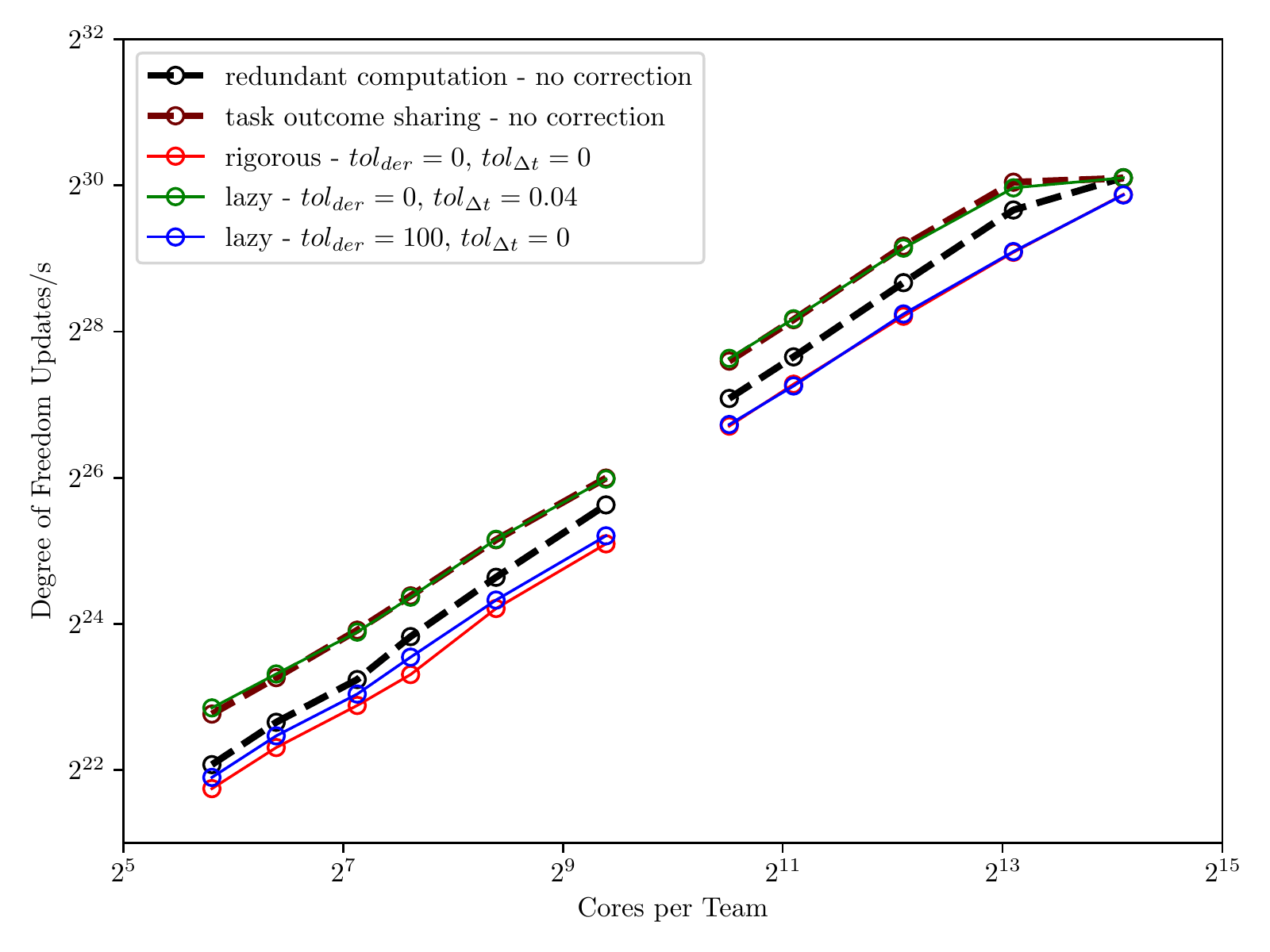}
\caption{Strong scaling of our benchmark on up to $35,088$ cores. A grid of $25^3$ cells is used for smaller core counts (left half, on $28$ ranks), while a grid of $79^3$ cells is employed for the larger setups (right half, on $731$ ranks).}
\label{fig:strong_scaling}
\end{figure}

We scale our benchmark on up to $35,088$ cores on SuperMUC-NG (Figure~\ref{fig:strong_scaling}), where we compare the performance of different configurations. Each configuration runs two teams\added[id=MB]{ (each on up to $17,544$ cores)}.
In the two baseline configurations, no errors are injected and our correction approach is disabled. 
Tasks are either computed redundantly (dashed black) or the two teams skip redundant computations using task outcome sharing~\cite{samfassISC} (dashed brown).
Besides, there are three different variants with error injection and correction: (1) a rigorous variant with high sensitivity and high redundancy (red), (2) a lazy variant with lower sensitivity but less redundancy (green) and (3) a lazy variant which --- in line with the results in Figure~\ref{fig:performance_tradeoff} --- we may assume to have a good performance-accuracy trade-off (blue). 
In all runs, we measure the performance for $100$ time steps. 
We inject (and correct) $10$ errors in each run with error correction. 
The error values and spatial positions are hardcoded to obtain a controlled and deterministic setup.
All experiments fix a number of MPI ranks for which our code results in a balanced domain decomposition of either a uniform grid with $25^3$ cells (left half in Figure~\ref{fig:strong_scaling}) or with $79^3$ cells (right half in Figure~\ref{fig:strong_scaling}). 
We then scale up the number of cores available to each rank.

The rigorous variant, as well as the blue lazy variant, have a lower performance than the baseline with redundant computation.
In both cases, error checking adds overhead for computing the error criteria, for transmitting outcomes and for waiting for their validation.
Compared to the red rigorous variant, only a selected subset of all cells is checked in the blue lazy one, which explains why the blue variant performs better for the small grid.
Yet, the non-uniform behavior of not validating all cells in the blue variant also creates load imbalances between ranks --- a bottleneck, especially at high rank counts.
The lazy green variant performs best, running at the full speed of the baseline that saves redundant computations, as error checking is applied to only few highly dubious cells.
No error correction overhead is visible for this variant.
All variants scale up similarly well, albeit small strong scaling effects at high core counts.

\todo{Brauchen wir hier noch ein takeaway?}

\section{Classification and discussion}
\label{section:evaluation-and-classification}

We next classify the properties of our approach, putting it into the context of related work where applicable.

\paragraph{Local vs.~global analysis}

Soft errors manifest in data corruption and can be found by running
a full simulation at least twice and comparing both results \cite{redundancy1, redundancy2, redundancy3}.
While both runs may be executed in parallel, the comparison requires an offline
or post-mortem analysis phase, once both redundant solutions are available.
Differing solutions either imply silent corruption or an error in the application itself.


Such a \emph{global} analysis challenges the supercomputer, as it introduces an
explicit synchronisation between the two redundant runs, and as the subsequent
comparison phase is very communication or I/O-heavy.
Such bursts stress critical components of the machine.
Instead, we adopt a \emph{local} approach where task outcomes are compared while the computation is running.
We thus avoid explicit synchronisation, and we spread out all comparisons
over the whole simulation time.

\paragraph{Check granularity}

Whenever we compare two redundantly computed solutions,
equality is to be interpreted in a numerical sense.
In a multi-threaded setting, the different orders of adding up individual
partial numerical results may result in two byte-wise disagreeing solutions.
\todo{Should we cite some of the reproducible numerics papers here? Maybe something from this special issue: https://journals.sagepub.com/toc/hpcc/34/5}
A similar argument holds for complex cache access patterns where data is put
from registers into memory and back.

The checks of redundantly computed results can happen at varying granularities: 
in principle, each individual floating point operation may be checked, i.e.~we
might compare data \emph{bit-wisely} subject to floating point precision.
In practice, this is often not feasible.
On the other hand, we might operate with \emph{global checksums} or hashes which
map the whole solution onto one or few characteristic values and compare these.
Our approach realises a bit-wise comparison in the tradition of the former
approach.
Yet, these checks are not automatically performed for all data all the time and
we hence meet our non-functional requirement that the resiliency routines may
not inacceptably increase the compute cost.

\paragraph{Redundancy level, check coverage and sensitivity}

Redundant computations are the method of choice to spot
errors and realise resiliency \cite{rMPI,redMPI}.
While our approach technically offers \emph{full redundancy}, i.e.~two
applications are in principle capable to run fully in parallel, the error criteria
checks reduce the logical redundancy level:
\replaced[id=PS]{Sharing}{The exchange of} task outcomes implies that we logically work with a 
\emph{partially redundant} setup where some data are assumed to be correct and
are not cross-checked.
There is no full coverage or validation of the outcomes.

We can only be \emph{weakly confident} that we spot silent data corruptions.
Strong errors resulting in NaNs or non-termination are always spotted by our
algorithm.
In return, our runtime data suggest that the sharing of confident outcomes
allows us to reduce the overhead cost of the redundant calculation
significantly.
Resiliency is not for free, but it does not increase the compute cost by
multiples of the baseline cost.
As we cache redundant computations only temporarily, the permanent memory
footprint also is not increased by multitudes.

Any error correction scheme must try to identify and correct potential soft errors as soon as possible to avoid error propagation. 
Ideally, an error is detected and corrected \emph{immediately}, i.e.,~before it actually has
the opportunity to affect any other numerical computations (immediate correction).
As we abandon the idea of global bit-wise comparisons, we deploy the
responsibility to identify errors immediately to the user code providing the
error criterion functions.

\paragraph{Recovery strategies}

Off-the-shelf solutions to recover from data corruption require simulations to
run at least three times such that the code can rely on a \emph{majority vote}
to identify which results are valid.
If one run is determined to be invalid, the whole simulation state is swapped,
i.e.~we continue after replicating a valid state.
This is a \emph{global recovery strategy}.

As we rely on cell-local error metrics, our approach realises a \emph{local
recovery strategy} where individual task outcomes are replaced if we spot an
error.
Furthermore, we rely solely on a \emph{redundancy level of two}.
Our error criteria replace the majority vote.
This allows us to operate without an expensive synchronisation after a time step
that brings a corrupted simulation back on track.
We also do not have to store complete checkpoints or hold them in memory.


\section{Conclusion and outlook}
\label{section:conclusion-and-outlook}

 We propose a task-based algorithmic framework which can recognise and
 correct\deleted[id=AR]{ing} soft errors.
 A clever combination of dubiosity metrics with task outcome sharing gives us an
 algorithm which exhibits almost the full robustness w.r.t.~errors of a fully
 replicated run without the runtime penalty.
 Checkpointing is completely eliminated although we have to keep task outcomes
 \replaced[id=AR]{in memory longer}{longer in memory} than the non-resilient baseline code and thus have a slightly
 increased memory footprint.
 While we have studied the impact for our ExaHyPE code and all of our work is
 open
 source,\footnote{\url{https://gitlab.lrz.de/exahype/ExaHyPE-Engine}}\footnote{\url{https://gitlab.lrz.de/hpcsoftware/teaMPI}} the paradigms and ideas are of relevance for a large set of explicit time stepping codes and, to the best of our knowledge, the only known alternative to standard
 checkpoint-restart or full replication.

\added[id=TW]{
 Natural follow-up work will combine the present ideas with algorithmic error
 correction techniques such as auto-correcting codes or limiter techniques.
 There are two further directions of future work worth exploring:
 On the one hand, a key ingredient of an efficient resilient realisation 
 is the fast evaluation of the error criteria.
 The evaluation does not necessarily have to be done on a compute node.
 With smart network devices, 
 task movement orchestration, labelling and merging may be offloaded into an
 intelligent network.
}

\added[id=TW]{
 On the other hand, the choice of proper tolerances is worth investigating.
 In our experiments, we chose fixed particular tolerance combinations and highlighted
 how they affect the runtime and sensitivity.
 At the same time, our experiments ``permitted'' errors to happen only in one
 step of the overall computation.
 This fact plus the potential task divergence for dubious task outcomes imply
 that soft errors still can sneak into a computation and pollute the long-term
 behaviour.
 However, our data suggests that harsh sensitivity thresholds can find errors in
 any task---a propagated error can formally be seen as a newly added error in a
 cell---and recover from them.
 It is thus reasonable to experiment with dynamic thresholds which are typically
 rather relaxed.
 If a system suspects that errors start to creep in, it is reasonable to
 increase the sensitivity and thus to recover also from long-term errors.
}

%
%

\bibliographystyle{IEEEtran}
\bibliography{IEEEabrv,references}

\begin{thebibliography}{10}
\providecommand{\url}[1]{#1}
\csname url@samestyle\endcsname
\providecommand{\newblock}{\relax}
\providecommand{\bibinfo}[2]{#2}
\providecommand{\BIBentrySTDinterwordspacing}{\spaceskip=0pt\relax}
\providecommand{\BIBentryALTinterwordstretchfactor}{4}
\providecommand{\BIBentryALTinterwordspacing}{\spaceskip=\fontdimen2\font plus
\BIBentryALTinterwordstretchfactor\fontdimen3\font minus
  \fontdimen4\font\relax}
\providecommand{\BIBforeignlanguage}[2]{{%
\expandafter\ifx\csname l@#1\endcsname\relax
\typeout{** WARNING: IEEEtran.bst: No hyphenation pattern has been}%
\typeout{** loaded for the language `#1'. Using the pattern for}%
\typeout{** the default language instead.}%
\else
\language=\csname l@#1\endcsname
\fi
#2}}
\providecommand{\BIBdecl}{\relax}
\BIBdecl

\bibitem{Dongarra:14:ApplMathExascaleComputing}
J.~Dongarra, J.~Hittinger, J.~Bell, L.~Chacon, R.~Falgout, M.~Heroux,
  P.~Hovland, E.~Ng, C.~Webster, and S.~Wild, ``Applied mathematics research
  for exascale computing,'' Tech. Rep., 2014.

\bibitem{Argonne:2014}
M.~Snir, R.~Wisniewski, J.~Abraham, S.~Adve, S.~Bagchi, P.~Balaji, J.~Belak,
  P.~Bose, F.~Cappello, B.~Carlson, A.~Chien, P.~Coteus, N.~Debardeleben,
  P.~Diniz, C.~Engelmann, M.~Erez, S.~Fazzari, A.~Geist, R.~Gupta, F.~Johnson,
  S.~Krishnamoorthy, S.~Leyffer, D.~Liberty, S.~Mitra, T.~Munson, R.~Schreiber,
  J.~Stearley, and E.~Hensbergen, ``Addressing failures in exascale
  computing,'' \emph{Int. J. High Perform. Comput. Appl.}, vol.~28, no.~2,
  2014.

\bibitem{Schroeder:2010:LargeScale}
B.~Schroeder and G.~A. Gibson, ``A large-scale study of failures in
  high-performance computing systems,'' \emph{{IEEE} Trans. Dependable Secur.
  Comput.}, vol.~7, no.~4, 2010.

\bibitem{Bronevetsky:2008}
G.~Bronevetsky and B.~R.~de Supinski, \emph{{Soft error vulnerability of
  iterative linear algebra methods}}, 2008.

\bibitem{Bronevetsky:2018}
G.~Bronevetsky, B.~R.~de Supinski, and M.~Schulz, ``A foundation for the
  accurate prediction of the soft error vulnerability of scientific
  applications,'' in \emph{IEEE Workshop on Sil. Err. in Logic}, 2018.

\bibitem{Austin:2015}
B.~Austin, E.~Roman, and X.~Li, ``Resilient matrix multiplication of
  hierarchical semi-separable matrices,'' in \emph{FTXS}.\hskip 1em plus 0.5em
  minus 0.4em\relax ACM, 2015.

\bibitem{Shantharam:2012}
M.~Shantharam, S.~Srinivasmurthy, and P.~Raghavan, ``Fault tolerant
  preconditioned conjugate gradient for sparse linear system solution,'' in
  \emph{ICS}.\hskip 1em plus 0.5em minus 0.4em\relax ACM, 2012.

\bibitem{sdc_multigrid}
M.~Altenbernd and D.~G\"{o}ddeke, ``Soft fault detection and correction for
  multigrid,'' \emph{Int. J. High Perf. Comp. Appl.}, vol.~32, no.~6, 2018.

\bibitem{Reinarz:2018}
A.~Reinarz, J.-M. Gallard, and M.~Bader, ``Influence of a-posteriori subcell
  limiting on fault frequency in higher-order dg schemes,'' in
  \emph{FTXS}.\hskip 1em plus 0.5em minus 0.4em\relax IEEE TCHPC, 2018.

\bibitem{Herault:2015}
T.~Herault and Y.~Robert, \emph{Fault-Tolerance Techniques for High-Performance
  Computing}.\hskip 1em plus 0.5em minus 0.4em\relax Springer, 2015.

\bibitem{Varela:2010}
M.~R. Varela, K.~B. Ferreira, and R.~Riesen, ``Fault-tolerance for exascale
  systems,'' in \emph{IEEE CLUSTER}, 2010.

\bibitem{redMPI}
D.~Fiala, F.~Mueller, C.~Engelmann, R.~Riesen, K.~Ferreira, and R.~Brightwell,
  ``Detection and correction of silent data corruption for large-scale
  high-performance computing,'' in \emph{SC}.\hskip 1em plus 0.5em minus
  0.4em\relax IEEE, 2012.

\bibitem{samfassISC}
P.~Samfass, T.~Weinzierl, B.~Hazelwood, and M.~Bader, ``Tea{MPI} -
  replication-based resilience without the (performance) pain,'' in \emph{ISC},
  2020.

\bibitem{lazy_shadowing}
R.~Melhem and T.~Znati, ``{Lazy Shadowing} - {A}n adaptive, power-aware,
  resiliency framework for extreme scale computing,'' Tech. Rep., oct 2019.

\bibitem{amt_resilience}
N.~Gupta, J.~R. Mayo, A.~S. Lemoine, and H.~Kaiser, ``Towards distributed
  software resilience in asynchronous many- task programming models,'' in
  \emph{FTXS}, 2020.

\bibitem{cpc_exahype}
A.~Reinarz, D.~Charrier, M.~Bader, L.~Bovard, M.~Dumbser, K.~Duru, F.~Fambri,
  A.-A. Gabriel, J.-M. Gallard, S.~Koeppel, L.~Krenz, L.~Rannabauer,
  L.~Rezzolla, P.~Samfass, M.~Tavelli, and T.~Weinzierl, ``{ExaHyPE}: An engine
  for parallel dynamically adaptive simulations of wave problems,''
  \emph{Comput. Phys. Commun.}, vol. 254, 2020.

\bibitem{Weinzierl:19:Peano}
T.~Weinzierl, ``{The Peano software - parallel, automaton-based, dynamically
  adaptive grid traversals},'' \emph{{ACM} Trans. Math. Softw.}, vol.~45,
  no.~2, 2015.

\bibitem{Dumbser:2006}
M.~Dumbser and M.~K{\"{a}}ser, ``{An arbitrary high-order discontinuous
  {G}alerkin method for elastic waves on unstructured meshes - II. The
  three-dimensional isotropic case},'' \emph{Geophy. J. Int.}, vol. 167, no.~1,
  2006.

\bibitem{Charrier2020}
D.~E. Charrier, B.~Hazelwood, and T.~Weinzierl, ``Enclave tasking for {DG}
  methods on dynamically adaptive meshes,'' \emph{{SIAM} J. Sci. Comput.},
  vol.~42, no.~3, 2020.

\bibitem{Charrier:19:EnergyAndDeepMemory}
D.~E. Charrier, B.~Hazelwood, E.~Tutlyaeva, M.~Bader, M.~Dumbser,
  A.~Kudryavtsev, A.~Moskovsky, and T.~Weinzierl, ``Studies on the energy and
  deep memory behaviour of a cache-oblivious, task-based hyperbolic {PDE}
  solver,'' \emph{Int. J. High Perform. Comput. Appl.}, vol.~33, no.~5, 2019.

\bibitem{krenzcloud}
L.~Krenz, L.~Rannabauer, and M.~Bader, ``A high-order discontinuous galerkin
  solver with dynamic adaptive mesh refinement to simulate cloud formation
  processes,'' in \emph{PPAM}.\hskip 1em plus 0.5em minus 0.4em\relax Springer,
  2019, pp. 311--323.

\bibitem{kelly}
J.~F. Kelly and F.~X. Giraldo, ``Continuous and discontinuous {G}alerkin
  methods for a scalable three-dimensional nonhydrostatic atmospheric model:
  Limited-area mode,'' \emph{J. Comput. Phys.}, vol. 231, no.~24, 2012.

\bibitem{continuations}
J.~Schuchart, P.~Samfass, C.~Niethammer, J.~Gracia, and G.~Bosilca,
  ``Callback-based completion notification using {MPI} continuations,''
  \emph{Parallel Comp.}, vol. 106, 2021.

\bibitem{redundancy1}
K.~Mohanram and N.~Touba, ``Cost-effective approach for reducing soft error
  failure rate in logic circuits,'' in \emph{ITC}, 2003.

\bibitem{redundancy2}
J.~Hu, F.~Li, V.~Degalahal, M.~Kandemir, N.~Vijaykrishnan, and M.~Irwin,
  ``Compiler-directed instruction duplication for soft error detection,'' in
  \emph{Design, Automation and Test in Europe}, 2005.

\bibitem{redundancy3}
A.~Messer, P.~Bernadat, G.~Fu, D.~Chen, Z.~Dimitrijevic, D.~Lie, D.~Mannaru,
  A.~Riska, and D.~Milojicic, ``Susceptibility of commodity systems and
  software to memory soft errors,'' \emph{{IEEE} Trans. Computers}, vol.~53,
  no.~12, 2004.

\bibitem{rMPI}
K.~B. Ferreira, J.~Stearley, J.~H.~L. III, R.~Oldfield, K.~T. Pedretti,
  R.~Brightwell, R.~Riesen, P.~G. Bridges, and D.~C. Arnold, ``Evaluating the
  viability of process replication reliability for exascale systems,'' in
  \emph{SC}.\hskip 1em plus 0.5em minus 0.4em\relax {ACM}, 2011.

\end{thebibliography}

\end{document}